\documentclass[aps, prb,twocolumn,superscriptaddress,showpacs,amsmath,amssymb, notitlepage]{revtex4-2}
\usepackage{amsfonts,hyperref}
\usepackage{float}
\usepackage{graphicx}
\usepackage{amsmath}
\usepackage{amssymb}

\usepackage{bm}

\usepackage{textcomp}

\begin{document}
\title{
Accurate computation of the energy variance for ground states and $\langle \langle \mathcal{L}^\dagger \mathcal{L} \rangle \rangle$ for steady states using iPEPS
}

\author{Emilio \surname{Cort\'es Estay}} \affiliation{Institute for Theoretical Physics, University of Amsterdam, Science Park 904, 1098 XH Amsterdam, The Netherlands}
\author{Naushad A. \surname{Kamar}} \affiliation{Institute for Theoretical Physics, University of Amsterdam, Science Park 904, 1098 XH Amsterdam, The Netherlands}
\author{Philippe \surname{Corboz}} \affiliation{Institute for Theoretical Physics, University of Amsterdam, Science Park 904, 1098 XH Amsterdam, The Netherlands}

\begin{abstract}
Infinite projected entangled-pair states (iPEPS) provide a powerful tensor network ansatz for two-dimensional quantum many-body systems in the thermodynamic limit. In this paper we introduce an approach to accurately compute the energy variance of an iPEPS, enabling systematic extrapolations of the ground-state energy to the exact zero-variance limit. It is based on the contraction of a large cell of tensors using the corner transfer matrix renormalization group (CTRMG) method, to evaluate the correlator between pairs of local Hamiltonian terms. We show that the accuracy of this approach is substantially higher than that of previous methods, and we demonstrate the usefulness of  variance extrapolation for the Heisenberg model, for a free fermionic model, and for the Shastry-Sutherland model. Finally, we apply the approach to compute $\langle \langle \mathcal{L}^\dagger \mathcal{L} \rangle \rangle$ for an open quantum system described by the Liouvillian $\mathcal{L}$, in order to assess the quality of the steady-state solution and to locate first-order phase transitions, using the dissipative quantum Ising model as an example.

\end{abstract}

\maketitle

\section{Introduction}
Tensor networks provide powerful numerical tools for studying strongly correlated quantum systems, particularly in regimes where Monte Carlo simulations fail due to the negative sign problem~\cite{troyer2005}. The best-known example are matrix product states which have revolutionized the study of (quasi) one-dimensional systems~\cite{white1992,schollwoeck2011}. Major progress has also been achieved with its higher dimensional generalization, known as projected entangled-pair states (PEPS)~\cite{verstraete2004,nishio2004}, which have become a state-of-the-art tool to study strongly correlated systems, ranging from frustrated magnets to strongly correlated electron systems; see, e.g., Refs.~\cite{corboz14_shastry, nataf16, liao17, niesen17, chen18, lee18, jahromi18, niesen18, yamaguchi18, kshetrimayum19b, chung19, ponsioen19, lee20, gauthe20, jimenez21, czarnik21, hasik21, shi22, liu22b, gauthe22, peschke22,hasik22, sinha22, ponsioen23b,  xu23b, weerda24,hasik24,schmoll24,zhang25}. A particular powerful variant is infinite PEPS (iPEPS)~\cite{jordan2008}, enabling to represent states directly in the 2D thermodynamic limit, with an accuracy that is controlled by the bond dimension $D$ of the tensors.

Obtaining an accurate estimate of the ground state energy in the exact infinite-$D$ limit is one of the central challenges in iPEPS calculations. 
A naive $1/D$ extrapolation is often neither controlled nor accurate, due to a typically non-monotonic dependence on $D$. Alternatives are provided by extrapolations as a function of truncation error within  the full-update imaginary time evolution algorithm~\cite{corboz16, zheng17,chung19}, or extrapolation in the inverse correlation length in the case of Lorentz-invariant critical states~\cite{corboz18,rader18,hasik21}. 
A common approach used in other variational methods~\cite{sorella01,iqbal13, viteritti25} is to extrapolate the energy as a function of the energy variance, $\mathrm{Var}(E) = \langle H^2 \rangle - \langle H \rangle^2$ which is zero for an exact eigenstate. However, the computation of $\langle H^2 \rangle$  involves evaluating non-local correlators between pairs of Hamiltonian terms, which is challenging with iPEPS. While different frameworks have been developed  to compute the variance~\cite{vanderstraeten16, corboz16b}, including example results for the Heisenberg model~\cite{vanderstraeten16}, these approaches have not yet found broad applications so far, partly because their implementation is challenging and their computational cost is high.

In this paper, we introduce a practical algorithm to compute the energy variance in a controlled and accurate way. It builds upon one of the standard contraction algorithms, the corner transfer matrix renormalization group (CTMRG) method~\cite{nishino1996,orus2009,corboz2011,corboz14_tJ}, which is used to evaluate all relevant pairs of Hamiltonian terms in a large cell. A similar idea was used in Ref.~\cite{arias24} for the computation of the dynamical structure factor, involving pairs of single-site operators.
A key feature of this large-cell (LC) CTMRG algorithm is that the required contraction dimension $\chi$ is substantially smaller than in previous schemes~\cite{corboz16b}, making it numerically tractable even for large bond dimensions. We test the approach for the $S=1/2$ antiferromagnetic Heisenberg model, for free fermions with a staggered chemical potential, and for the Shastry-Sutherland model (see also Ref.~\cite{corboz25} for additional results), and demonstrate that the variance provides a useful quantity to extrapolate the iPEPS energy.

Finally, we show how this technique can  be used in the context of open quantum systems to compute $\epsilon=\langle \langle \mathcal{L^\dagger L} \rangle\rangle$, with  $\mathcal{L}$ the Liouvillian, 
in order to assess the closeness of the iPEPS to the exact steady state characterized by $\epsilon=0$. We demonstrate how this quantity can be used to locate first-order phase transitions, using the dissipative transverse-field Ising model as an example, thus providing a practical tool to study steady-state phase diagrams with iPEPS.

\section{Method}
\label{sec:method}
\subsection{Introduction to iPEPS and CTRMG}
\begin{figure}[t]
  \centering
  \includegraphics[width=1\linewidth]{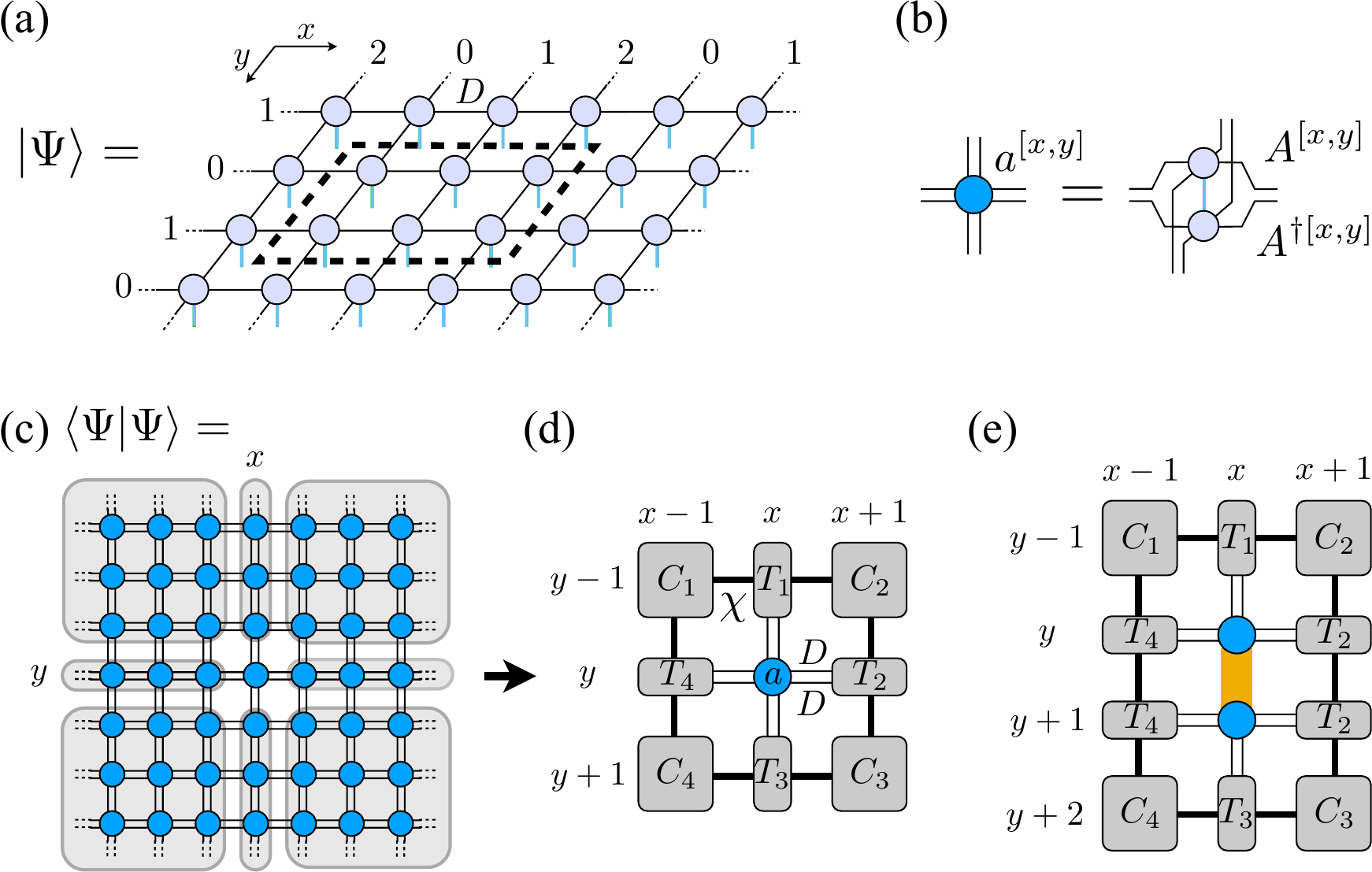}
  \caption{(a) iPEPS with a $3 \times 2$ unit cell. (b) Double-layer tensor $a$ at coordinate $[x,y]$, defined as the contraction of the iPEPS tensor $A$ with its conjugate $A^\dagger$ along the physical leg.
  (c)  Norm of the state, represented by an infinite square-lattice tensor network of the $a$ tensors. 
(d) In CTMRG the tensor network surrounding a central site is effectively encoded in four corner ($C$) and edge ($T$) environment tensors. The coordinates $[x,y]$ denote the relative position in the unit cell. (e) Example of the relevant tensor-network diagram for the expectation value of a nearest-neighbor Hamiltonian term in vertical direction (orange rectangle), which is sandwiched between the physical legs of the $A$ and $A^\dagger$ tensors.}
  \label{fig:ctm}
\end{figure}

An iPEPS is a variational tensor-network ansatz for representing 2D states directly in the thermodynamic limit~\cite{jordan2008}. It consists of a unit cell of rank-5 tensors $A^{[x,y]}$ repeated on a square lattice, where each tensor has 4 auxiliary legs with bond dimension $D$ connecting neighboring tensors and one physical leg with the dimension $d$, corresponding to the local Hilbert space of a lattice site; see Fig.~\ref{fig:ctm}(a). The coordinates $[x,y]$ indicate the position of a tensor within the unit cell.  
Combining each tensor $A^{[x,y]}$ with its conjugate $A^{\dagger[x,y]}$ into a new tensor $a^{[x,y]}$ (Fig.~\ref{fig:ctm}(b)) results in a square-lattice tensor network shown in Fig.~\ref{fig:ctm}(c), representing the norm of the wave function.

To contract the infinite tensor network, we use the CTMRG method~\cite{nishino1996, orus2009}, generalized to large unit cells~\cite{corboz2011, corboz14_tJ}. In CTMRG the infinite tensor network surrounding a central site is effectively encoded in an environment consisting of four corner and four edge tensors; see Fig.~\ref{fig:ctm}(d). In case of a multi-site unit cell, separate environment tensors are computed for each position in the unit cell. These environment tensors are obtained by an iterative procedure, in which columns and rows of the unit cell are absorbed into the environment tensors, followed by a renormalization step in which the enlarged bond dimension is truncated based on a pair of projectors; see Refs.~\cite{corboz2011, corboz14_tJ} for details. A growth step in which a column of tensors is absorbed into the left boundary is called a left move, and one similarly defines a right, top, and bottom move.
The accuracy of the contraction is systematically controlled by the boundary bond dimension $\chi$ of the environment tensors. 

Local expectation values can be obtained by sandwiching the operators between the physical legs of the iPEPS tensors and embedding them in the corresponding environment. Figure~\ref{fig:ctm}(e) shows an example of the diagram for one of the vertical Hamiltonian terms in the unit cell. Since the iPEPS is not normalized, the result needs to be divided by the norm represented by the same network without the operator.

In order to obtain the tensor network representation of the ground state of a given Hamiltonian, the optimal variational parameters of the tensors need to be determined. This has traditionally been done by performing an imaginary-time evolution based on the simple update~\cite{jiang2008} or full update~\cite{jordan2008,phien15} method. More recently, more accurate optimizations have been achieved by directly minimizing the variational energy~\cite{corboz16b, vanderstraeten16,liao19}, in particular based on automatic differentiation~\cite{liao19}. In this work, we use the latter approach to determine the optimal variational parameters.

\subsection{Previous methods to compute the energy variance}
In this section we briefly discuss previous and related frameworks to compute the energy variance with iPEPS. For simplicity, we consider a translationally invariant iPEPS with a single-site unit cell representing the ground state of a nearest-neighbor Hamiltonian $H=\sum_b H_b$. The energy variance is defined as
\begin{equation}
\mathrm{Var}(E) = \langle H^2 \rangle - \langle H \rangle^2 = \sum_{b,b'} \left( \langle H_b H_{b'} \rangle - \langle H_b \rangle \langle H_{b'} \rangle \right)
\end{equation}
and the variance density, relevant for iPEPS, corresponds to the variance per lattice site. The second term can be easily obtained from  local expectation values (cf. Fig.~\ref{fig:ctm}(e)). The first term, however, is more challenging. 
By exploiting translational invariance, we can reduce the double infinite sum to a single one, by placing the second term at a certain reference position, e.g. the center of the system. 
Hence, we need to evaluate  $ \langle (\sum_b H_b) H_0 \rangle$, with $H_0$ a Hamiltonian term at the center.
In the case where the state is not rotationally symmetric, we have to separately evaluate  the Hamiltonian term on the horizontal and on the vertical bond in the center, or more generally, on all non-equivalent bonds in the unit cell in the case of a non-translationally invariant iPEPS.

Different frameworks to perform summations of Hamiltonian terms for iPEPS have been developed in the past. In Ref.~\cite{corboz16b} a CTMRG-based method was introduced to encode summed Hamiltonian contributions in so-called $H$-environment tensors, similar to the usual corner and edge tensors, but including the summed Hamiltonian terms. Based on these $H$-environment tensors, the variance can be computed by sandwiching an additional Hamiltonian term in the center of the system. An alternative summation scheme is provided by so-called channel environments, introduced in Ref.~\cite{vanderstraeten16}, with example variance computations presented for the Heisenberg model. 

One shortcoming of these approaches is that the projectors used in the renormalization procedure do not take into account the presence of the Hamiltonian terms, leading to a relatively slow convergence as a function of the contraction dimension $\chi$. While this is not a problem at small bond dimensions, for large bond dimensions the required $\chi$ becomes prohibitively large, as we will show in the results section. The same issue of slow $\chi$-convergence was observed for the structure-factor computation~\cite{ponsioen23,vanderstraeten22}, i.e., the summed two-point spin-spin correlator. It was shown that speed of convergence can be substantially improved by adding a correction to the projectors~\cite{ponsioen23} to account for the presence for the summed spin operators in the CTMRG environments. This scheme is also closely related to the generating-function approach to compute summed correlation functions~\cite{tu21,tu24}. Applying this idea to the case of a summed correlation function of Hamiltonian terms is more challenging; for this reason we follow a different route in this work, inspired by Ref. \cite{arias24}.

\subsection{Variance computation using large-cell CTMRG contraction}
\label{sec:method2}
In Ref.~\cite{arias24} an accurate scheme for computing two-point spin-spin correlation functions  was introduced to evaluate the (dynamical) spin structure factor, based on the contraction of a large cell using CTMRG. Here we adopt and extend this idea to evaluate correlations between Hamiltonian terms, thereby enabling to compute the energy variance. We first consider the case of a translationally invariant iPEPS with a single tensor $A$, representing a state  $| \Psi \rangle$, together with converged CTMRG environment tensors, and later explain how to extend the scheme to  multi-site unit-cell case.

\begin{figure}[tb]
  \centering
  \includegraphics[width=1\linewidth]{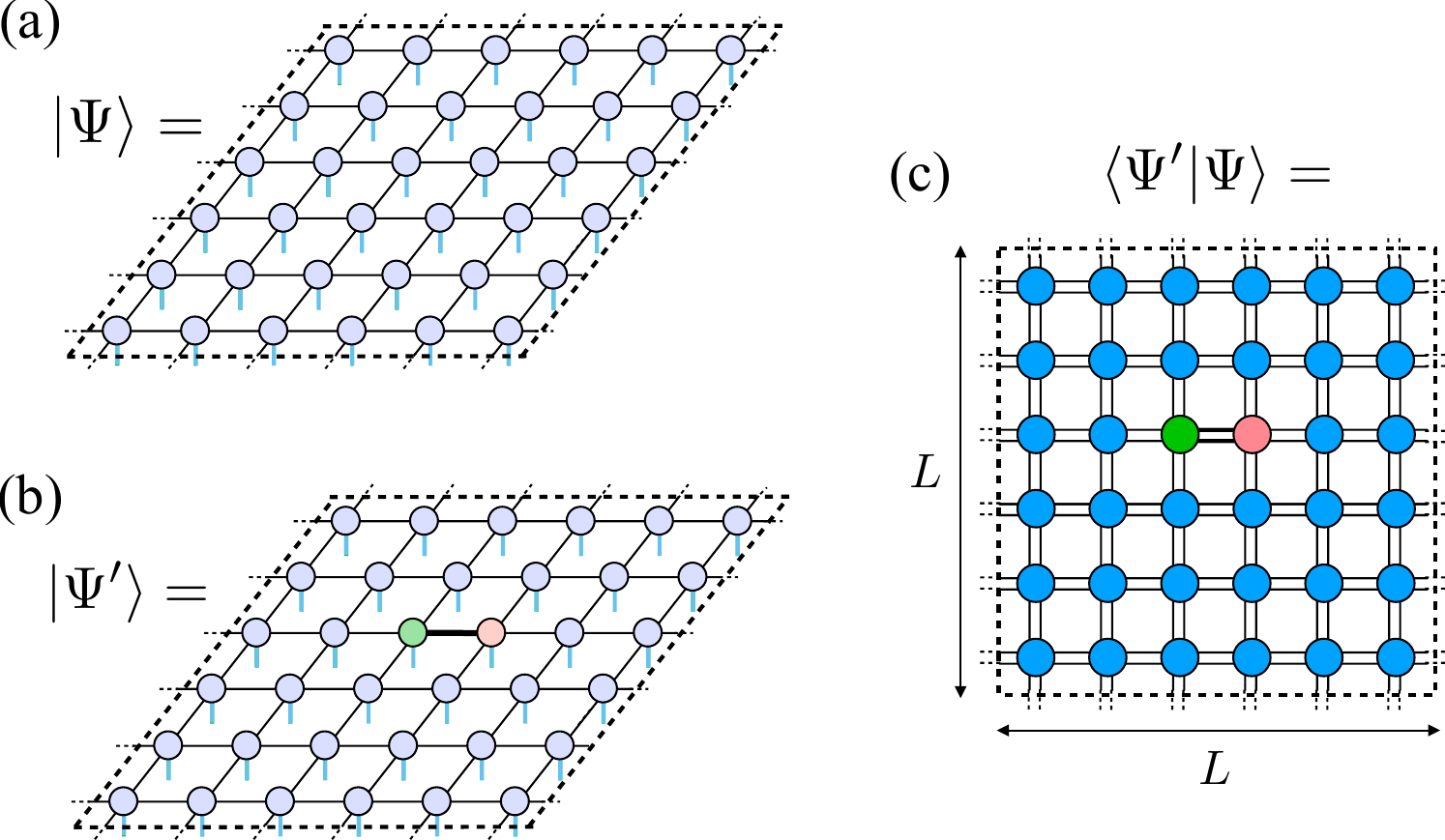}
  \caption{
  (a) Translationally invariant iPEPS with the unit cell  enlarged to size $L\times L$. (b) iPEPS wavefunction from (a) with a  Hamiltonian term $H_0$ applied in the center of the cell, resulting in $| \Psi' \rangle$. (c) The overlap  $\langle \Psi | \Psi' \rangle$, which we contract using CTMRG, starting from the central columns.
 }
  \label{fig:lce}
\end{figure}

The LC-CTMRG approach consists of the following steps:
\begin{enumerate}
\item
We first enlarge the unit cell to size $L \times L$ by  copying the tensor $A$ to all positions in the cell; see Fig.~\ref{fig:lce}(a). We also initialize the CTMRG environment tensors at each coordinate with the previously computed environment tensors.  

\item
We apply a Hamiltonian term to the iPEPS in the center of the unit cell, resulting in two new tensors $B$ and $C$ with enlarged bond dimension between them  (Fig.~\ref{fig:lce}(b)), representing a new state $| \Psi' \rangle = H_{0} | \Psi \rangle $. The correlator between a Hamiltonian term $H_b$ at a certain position and the one in the center, $H_0$ can be rewritten as
\begin{equation}
\frac{\langle \Psi | H_b H_0 | \Psi \rangle}{\langle \Psi | \Psi \rangle}  = \frac{\langle \Psi | H_b | \Psi' \rangle}{\langle \Psi | \Psi' \rangle}       \frac{\langle \Psi | \Psi' \rangle}{\langle \Psi | \Psi 
\rangle} 
\end{equation}
The second term on the right-hand side simply corresponds the local bond energy at the center, $\langle H_0 \rangle$, whereas the first term can be computed based on CTMRG as follows.

\item
We use CTMRG  to contract the tensor network representing the overlap $\langle \Psi | \Psi' \rangle$. For a horizontal Hamiltonian term, we start from the center and perform left moves up to the right boundary and then, again starting from the center, right moves up to the left boundary, followed by analogous top and bottom moves (see below for further details). This sweeping procedure can be repeated several times to improve convergence of the environment tensors. In practice, we found that two  sweeps are sufficient.
\item
With the converged CTMRG environment tensors, we can now readily evaluate $\frac{\langle \Psi | H_b | \Psi' \rangle}{\langle \Psi | \Psi' \rangle}$ for each position of the Hamiltonian term $H_b$, in a similar fashion to a standard local expectation value.
\end{enumerate}

This approach comes with several advantages. First, the core of the algorithm is based on the large-unit-cell CTMRG algorithm~\cite{corboz2011, corboz14_tJ}; i.e., existing codes can be easily adapted to compute the variance (in contrast to the previous algorithms~\cite{corboz16b, vanderstraeten16} which are much more complex). Second, since the CTMRG scheme is applied to the computation of the overlap   $\langle \Psi | \Psi' \rangle$, the CTMRG projectors used in the renormalization step naturally take the presence of $H_0$ into account, leading to a substantially improved accuracy, as we will show in the results section. Third, since each term is computed separately, the convergence of individual terms and behavior as a function of distance can be effectively monitored.

We end this section with a few more technical details. The precise sequence of update moves for the contraction with a horizontal Hamiltonian term in the center is as follows. Let $m=L/2$ be the middle position. We first perform a right move at position $x=m+1$ to initialize the  environment tensors on the right side ($C_2$, $T_2$, $C_3$) at position $m+1$ (for all $y$ positions), such that they take into account the updated bond. We then perform left moves from position $x=m-1$ up to $L-1$. We start from $m-1$ rather than $m$ to improve the left boundary tensors ($C_1$, $T_4$, $C_4$) in the vicinity of the new $B$ and $C$ tensors. By stopping the sweep at $L-1$ rather than $L$ we make sure that the correlations of the center Hamiltonian are not carried over to the neighboring cell. We then proceed with right moves from position $x=m+2$ down to 2, followed by top moves from position $y=m-1$ up to $L-1$ (for all $x$) and bottom moves from $y=m+1$ down to 2. In practice, we perform a second sweep (without the initialization step) to improve convergence. For a vertical Hamiltonian term, we proceed in a similar fashion, just with all moves and coordinates rotated by 90 degrees.

The computational cost of the approach scales as $O(L^2)$ times the scaling of a single CTMRG renormalization step, $O(D^4 \chi^3)$ (using an iterative solver to compute the CTMRG projectors). The CTMRG steps at the center of the cell incur an additional prefactor due to the enlarged bond dimension $D' = kD > D$ of the center bond, but there are only $O(1)$ such steps. Hence, compared to a standard evaluation of the energy, which requires $\approx L/2$ CTMRG steps, the large-cell contraction is roughly $2(L+k)$ times more expensive, using two sweeps, from the center to the boundaries, as discussed in the previous paragraph. The evaluation of all local expectation values is $L^2$ times more expensive, however, this part is typically fast compared to the computation of the projectors.

In the case of a ground-state iPEPS with a larger unit-cell size as a starting point, the computation is repeated for each inequivalent Hamiltonian term position in the unit cell of the state, which can easily be parallelized for efficiency. The cell size for the variance evaluation is then chosen to be a multiple of the initial unit-cell size.


\section{Results}
\subsection{Heisenberg model}
We first benchmark the approach for the $S=1/2$ antiferromagnetic Heisenberg model. We use an iPEPS with two tensors, one for each sublattice, with implemented U(1) symmetry~\cite{singh2010,bauer2011}.

\begin{figure}[t]
  \centering
  \includegraphics[width=1\linewidth]{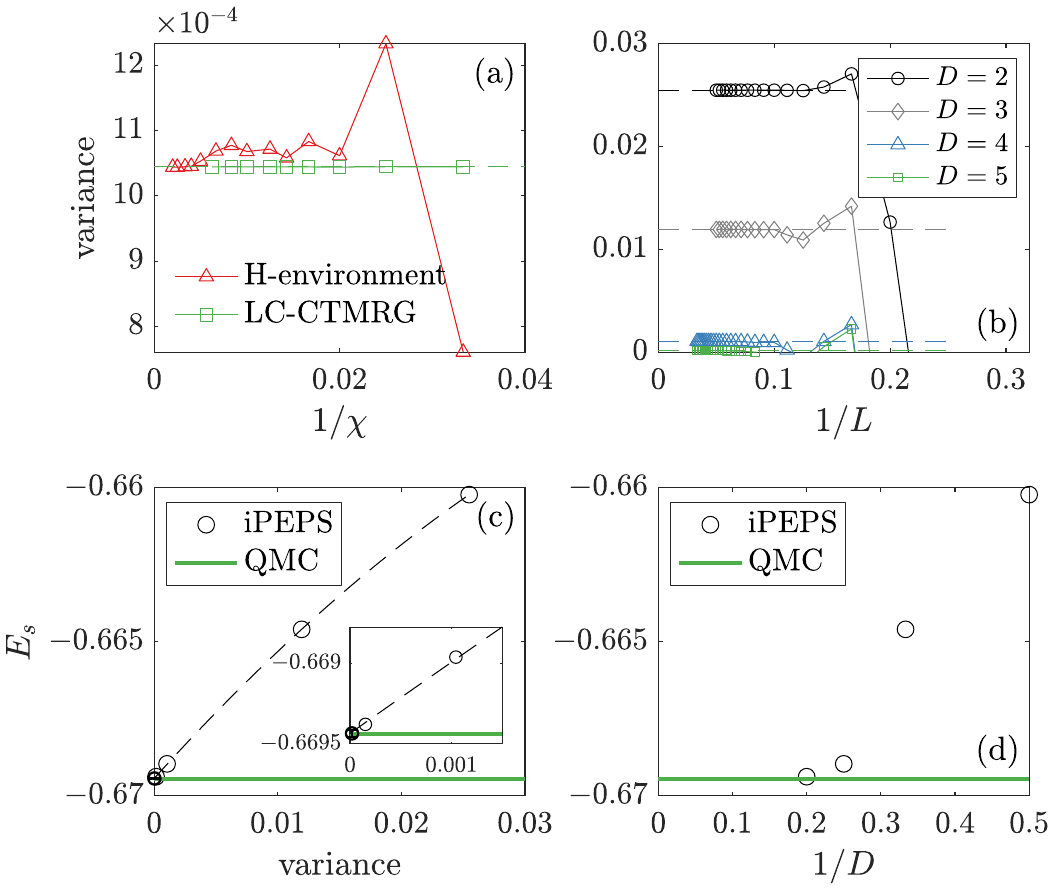}
  \caption{(a) Convergence of the variance as a function of inverse $\chi$ for $D=4$ and large $L=30$, comparing the previous $H$-environment approach from Ref.~\cite{corboz16b} with the LC-CTMRG method introduced in this work. (b) Convergence of the variance with cell size $L$ for different values of $D$. (c) Energy as a function of the variance. Using a second-order polynomial extrapolation yields a value compatible with the QMC result. The inset shows a zoomed-in view. (d) Energy as a function of inverse bond dimension, showing a non-monotonic behavior, which makes it difficult to extrapolate to the infinite $D$ limit. }
  \label{fig:heis}
\end{figure}

Figure~\ref{fig:heis}(a) shows the convergence of the variance as a function of inverse $\chi$ for $D=4$, comparing the current approach with the previous $H$-environment method from Ref.~\cite{corboz16b}. The latter displays slow convergence as a function of $\chi$, with values of $\chi>200$ required for a converged result. In stark contrast, the LC-CTMRG approach yields an accurate estimate already at much smaller values, around $\chi = 30$, providing clear evidence of its increased precision.

In Fig.~\ref{fig:heis}(b) we show the convergence of the variance as a function of the cell size $L$ for different values of $D$. The larger $D$ is, the larger the required cell size to reach convergence. This can be naturally understood from the fact that the finite bond dimension $D$ introduces an effective correlation length, which grows as $D$ increases and reproduces the infinite correlation length of the gapless ground state in the infinite-$D$ limit~\cite{corboz18,rader18}. The longer the correlation length, the larger the required cell size  to capture all relevant terms contributing to the variance~\footnote{We note that even in the presence of long-range order, the correlation length extracted from the connected correlation function remains finite (at finite $D$), because the iPEPS explicitly breaks the symmetry.}. We note that studying the convergence as a function of $L$ does not require separate calculations for each $L$; in practice, one performs a single calculation at a sufficiently large $L_{\mathrm{max}}$, from which one can directly extract the data for smaller $L$, since all relevant terms are computed individually.

Finally, the energy as a function of the variance is plotted in Fig.~\ref{fig:heis}(c). Using a second-order polynomial fit based on the $D=2\ldots5$ data, we obtain an energy in the zero-variance limit of $E_s=-0.669436(21)$, in agreement with the quantum Monte Carlo result $E_{QMC}=-0.6694421(4)$ from Ref.~\cite{sandvik2010}, illustrating the usefulness of variance-based energy extrapolations with iPEPS. For comparison, the data is also plotted as a function of inverse bond dimension in Fig.~\ref{fig:heis}(d), which is difficult to extrapolate due to its non-monotonic behavior.

\subsection{Free fermions with staggered chemical potential}
While for gapless systems the energy could also be extrapolated based on finite-correlation-length scaling~\cite{corboz18,rader18,hasik21}, this is not the case for gapped systems. To test the variance extrapolation in the latter case, we consider a model of free fermions on a square lattice, where a gap is induced by a staggered chemical potential, given by the Hamiltonian
\begin{equation}\label{eq:hub}
\hat H=-t \sum_{\langle i,j, \sigma\rangle}  \left(\hat c_{i\sigma}^\dagger \hat c_{j\sigma}  + H.c.\right) +\Delta \sum\limits_{i \in {\cal A}} \hat n_{i} -  \Delta \sum\limits_{j \in {\cal B}} \hat n_{j},
\end{equation}
where $\hat c^\dagger_{i\sigma}$ ($\hat c_{i\sigma}$) creates (annihilates) an electron with spin $\sigma=\left\{\uparrow, \downarrow\right\}$ on site $i$,  $\hat n_{i}=\sum_\sigma \hat c^\dagger_{i\sigma}\hat c_{i\sigma}$ is the number operator, $t=1$ the hopping amplitude, and $\Delta$ the magnitude of the staggered chemical potential, which has opposite signs on the two sublattices $\cal A$ and $\cal B$.

\begin{figure}[tb]
  \centering
  \includegraphics[width=1\linewidth]{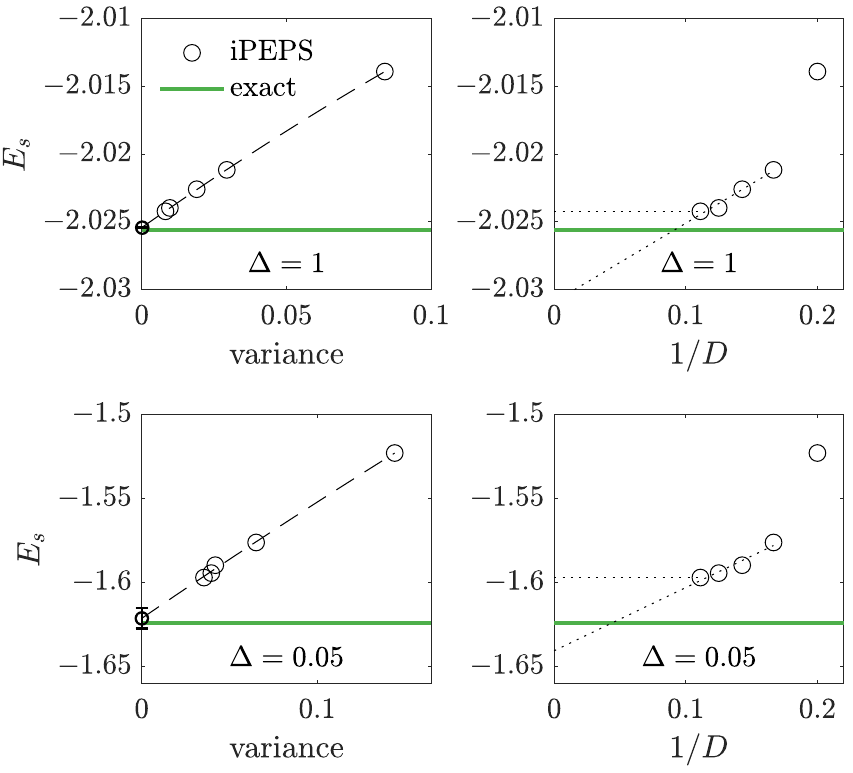}
  \caption{Energy per site as a function of the variance (left panels) and inverse bond dimension $1/D$ (right panels), for two different values of $\Delta$. The extrapolated results based on the variance are in agreement with the exact results, indicated by the green lines. The dotted lines in the right panels are a guide to the eye. 
 }
  \label{fig:fermions}
\end{figure}

Figure~\ref{fig:fermions} shows the energy per site as a function of the variance and  $1/D$ for two different values of $\Delta$ and $D$ in the range $[5,9]$, computed using sufficiently large $L$ and $\chi$ for each value of $D$ for accurate estimates.
Extrapolating the energy to the zero-variance limit using a second-order polynomial yields an estimate of the ground-state energy in agreement with the exact result, indicated by the green lines. In contrast, it is not obvious how to best extrapolate the data as a function of $1/D$. A common approach is to consider the range of energies spanned by the largest $D$ value (upper bound) and the linearly extrapolated value. The latter usually provides a lower bound, since the energy typically converges faster than linearly. While the exact result lies within this interval, the associated uncertainty is large. 

\begin{figure}[tb]
  \centering
  \includegraphics[width=1\linewidth]{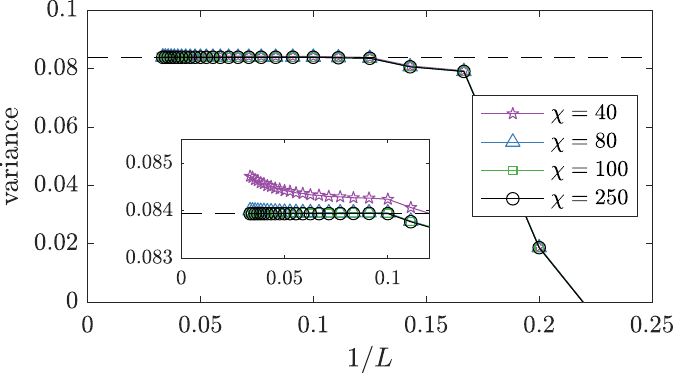}
  \caption{Convergence of the variance as a function of cell size $L$, for different values of $\chi$. A small value of $\chi$ can lead to an artificial drift of the variance value with increasing cell size, as can be seen in the $\chi=40$ data.}
  \label{fig:fermionsconv}
\end{figure}

We note that also in this case a relatively small $\chi$ is sufficient to obtain an accurate estimate of the variance. One subtlety to be careful about is a potential drift of the value for large cell sizes and small $\chi$, as illustrated in Fig.~\ref{fig:fermionsconv}. On the full scale, already $\chi=40$ provides a good estimate of the variance for a cell size of $L=10$, with a relative error of only 0.4\%. However, when zooming in (inset), one can observe a drift of the $\chi=40$ curve with increasing $L$, such that at large $L$ the result becomes less accurate. This is due to the fact that small $\chi$ can lead to inaccuracies at long distances, adding noise to the variance. The connected correlator of  Hamiltonian terms, $\langle H(r) H_0 \rangle - \langle H(r) \rangle \langle H_0 \rangle$, should vanish in the large-distance limit; however, contraction errors can lead to imperfect cancellation at large distances, resulting in an additive contribution with increasing $L$. Hence, the convergence as a function of $L$ has to be monitored and checked for drifts, especially when relatively small values of $\chi$ are used.

\subsection{The Shastry-Sutherland model}

The Shastry-Sutherland model~\cite{Shastry81,Miyahara03} is an effective model of the layered material SrCu$_2$(BO$_3$)$_2$, and has been the subject of intense studies over the past decades~\cite{Koga00,Laeuchli02,Corboz13_shastry,yang22,wang22,keles22,viteritti25,xi23,qian24,liu24,maity25}. It is described by a frustrated  $S=1/2$ spin model on an orthogonal dimer lattice, given by the Hamiltonian
\begin{equation}
H=J\sum_{\langle i,j \rangle}\bm S_{i}\cdot \bm S_{j}+J'\sum_{\langle \langle i,j \rangle\rangle}\bm S_{i}\cdot \bm S_{j}
\end{equation}
with $J$ and $J'$ the intra- and interdimer coupling strengths, respectively. For small values of $J'/J$ the ground state is an exact product of dimer singlets~\cite{Shastry81}. In the $J'/J \rightarrow \infty $ limit, the model reduces to the square-lattice Heisenberg model with an AF ground state. For intermediate $J'/J$ a valence-bond solid phase is realized, with strong bonds forming around plaquettes~\cite{Koga00,Laeuchli02,Corboz13_shastry}. More recently, there has been growing consensus regarding the existence of an additional narrow quantum spin liquid phase between the plaquette and antiferromagnetic phases~\cite{yang22,wang22,keles22,viteritti25,maity25,corboz25}.

The variance calculation based on LC-CTMRG was already applied in Ref.~\cite{corboz25} to extrapolate the energies of the plaquette and quantum spin liquid phase to determine the phase transition. A benchmark comparison between different methods at $J'/J=0.8$ showed that iPEPS with bond dimensions up to $D=10$, combined with the variance extrapolation, provides the most accurate energy estimate of the ground state in the thermodynamic limit, $E_s=-0.44896(2)$, as compared to the estimate from neural network quantum states~\cite{viteritti25}, $E_s=-0.4486(2)$, or from density matrix renormalization group, $E_s=-0.44837(30)$, \cite{yang22}.

\begin{figure}[tb]
  \centering
  \includegraphics[width=1\linewidth]{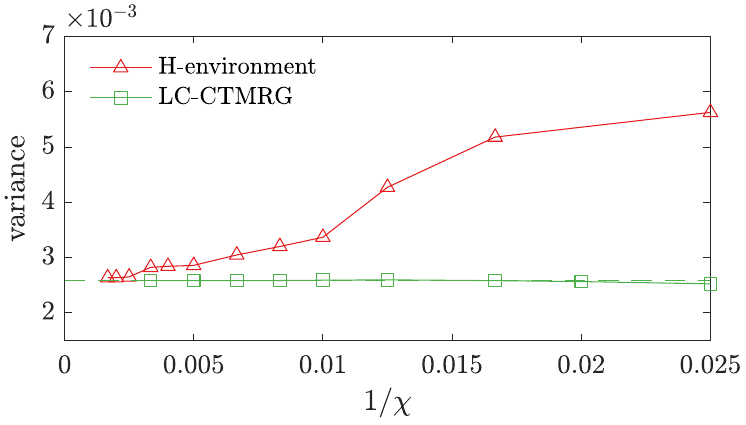}
  \caption{Convergence of the variance as a function of inverse $\chi$ for the Shastry-Sutherland model with $J'/J=0.8$, \mbox{$D=5$}, comparing the previous $H$-environment approach with the LC-CTMRG method.
 }
  \label{fig:ssm}
\end{figure}

While we refer to Ref.~\cite{corboz25} for the main data and extrapolations, here we present another comparison between the previous $H$-environment method~\cite{corboz16b} with the LC-CTMRG approach. Figure~\ref{fig:ssm} shows the energy variance for $D=5$ for $J'/J=0.8$ as a function of $1/\chi$. As already observed for the Heisenberg model, LC-CTMRG converges much more rapidly than the $H$-environment approach, which is still not fully converged at the largest bond dimension $\chi=300$ tested. This makes it nearly impossible to extract the variance in a reliable way at even larger bond dimensions with the $H$-environment scheme, in contrast to the LC-CTMRG approach.


\section{Open quantum systems}
The LC-CTMRG approach also finds useful application for steady states of open quantum systems. Consider an open quantum system described by a Markovian master equation in Lindblad form
\begin{equation}
	\dot{\rho} = \mathcal{L} [ \rho] = -i\left[H,\rho \right] + \sum_{\mu}\left( L_\mu \rho L_\mu^\dagger - \frac{1}{2} \{ L_\mu^\dagger L_\mu,  \rho \} \right), 
	\label{master0}
\end{equation}
with $\mathcal{L}$ the Liouvillian superoperator and $L_\mu$ local Lindblad operators. For a two-dimensional system, iPEPS can be used to represent the vectorized density matrix, $ | \rho\rangle\rangle$, and the steady state of the system can be obtained via a long-time evolution
\begin{equation}
| \rho_s\rangle\rangle = \lim_{T\rightarrow \infty} e^{T\mathcal{L}} | \rho_0\rangle\rangle,
\end{equation}
using a Trotter-Suzuki decomposition of the evolution operator, together with simple- or full-update schemes to perform the time evolution~\cite{kshetrimayum17,czarnik19,mckeever21}.

The steady state fulfills $\mathcal{L} | \rho_s\rangle\rangle = 0$ in the exact infinite bond dimension limit; hence the expectation value
\begin{equation}
\epsilon = \langle \langle \rho_s| \mathcal{L}^\dagger  \mathcal{L} | \rho_s \rangle\rangle 
\end{equation}
can be used to assess the closeness to the true steady state. This expectation value has the same structure as $\langle H^2 \rangle$ in the energy-variance computation; therefore, we can use the LC-CTMRG approach (or previous frameworks~\cite{vanderstraeten16,corboz16b}) to accurately compute $\epsilon$ (per site), enabling to assess the quality of the tensor-network representation of the steady state.

The quantity $\epsilon$ can also be used to locate a first-order phase transition, inspired by similar ideas from ground-state computations. In the vicinity of a first-order phase transition, hysteresis  can be observed; i.e., a state in one phase remains metastable even slightly across the phase transition. In ground-state calculations, this can be exploited by initializing simulations in both phases and tuning the coupling parameter across the  transition from both sides. The first-order phase transition can then be found by determining the point where the energies of the two states intersect. Here we apply the same idea to open quantum systems, but instead of considering the energy, we use $\epsilon$ as the diagnostic quantity.

We test this approach for the dissipative quantum Ising model~\cite{lee11,marcuzzi14, weimer15a,weimer15b}, defined by the Hamiltonian and jump operators
\begin{equation}
H = \frac{V}{4} \sum_{\langle i,j \rangle} \sigma^z_i \sigma^z_j + \frac{h_x}{2} \sum_i \sigma^x_i, \quad L_\mu = \sqrt{\gamma} \sigma^-_\mu
\end{equation}
where V is the magnitude of the coupling, $h_x$ the transverse magnetic field strength, and $\gamma$ controls the strength of the dissipation. We consider the same parameters as in previous studies~\cite{weimer15a,kshetrimayum17}, $V=5\gamma$, $\gamma=0.1$, where a first order phase transition was found around $h_x/\gamma\sim 6\ldots7 $ between two phases with different densities of up spins.

\begin{figure}[t]
  \centering
  \includegraphics[width=1\linewidth]{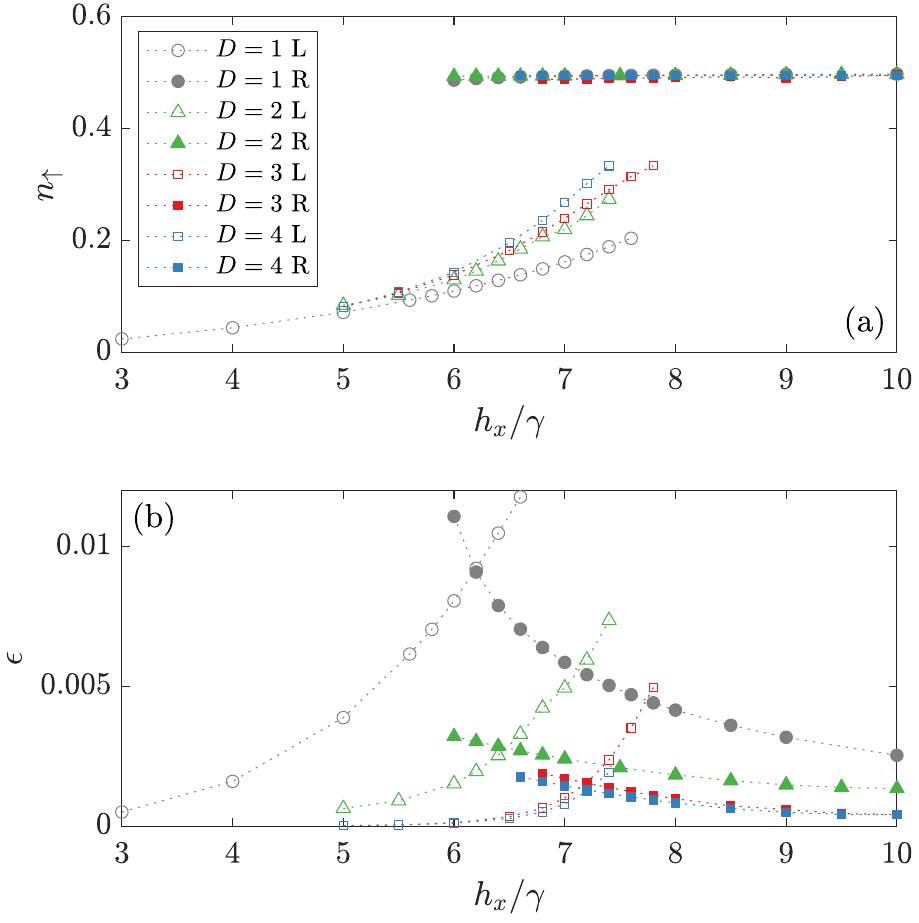}
  \caption{(a) Up-spin density $n_\uparrow=\langle(1+\sigma_z)/2\rangle$ as a function of $h_x/\gamma$ for $V=5$ for different values of $D$. Open and full symbols correspond to values from simulations initialized in the left and right phase, respectively. The states remain metastable across the phase transition due to its first-order nature. (b) Expectation value  $\langle \langle \cal{L^\dagger L} \rangle \rangle$ indicating the closeness to the true steady-state, measured in both phases. We take the intersection of each pair of curves as an estimate of the location of the first-order transition. The intersection at the largest bond dimension $D=4$ is found around $h_x/\gamma=7.2$.}
  \label{fig:openising}
\end{figure}

We perform iPEPS simulations of the steady state of this system using a combination of simple update~\cite{kshetrimayum17} and full update~\cite{czarnik19}, with the magnitude of $\epsilon$ as a criterion for the quality of the state, computed for a cell size of $L=12$~\footnote{Here we have used a cell size $L=12$ based on which the connected correlator is sufficiently converged. However, we note that unlike the connected correlator, $\langle\langle {\mathcal L^\dagger}(r) {\mathcal L}_0 \rangle\rangle$ does not vanish at large distances, because typically  $\langle\langle \mathcal{L}\rangle\rangle \neq 0$ for the approximate steady-state. }. We start from different initial random iPEPS on both sides of the phase transition and select the ones with lowest $\epsilon$. Starting from these reference states in both phases, we move across the phase transition from both sides. Throughout the evolution, we monitor the value of $\epsilon$ and keep the state with lowest $\epsilon$. For each value of $D$, the transition is determined from the intersection in $\epsilon$ of the two phases, as shown in Fig.~\ref{fig:openising}. Based on this procedure, we find that the location of the transition moves from $h_x/\gamma = 6.2$ for $D=1$ to a larger value $h_x/\gamma = 6.5$ for $D=2$, and then to $h_x/\gamma = 7.2$ for $D=3$ and $D=4$, providing our best estimate.

We note that it would be interesting to perform the optimization of the steady state directly by minimizing $\langle \langle \cal{L^\dagger L} \rangle \rangle$. While this is technically possible via automatic differentiation similarly as in ground state optimization~\cite{liao19}, the computational cost would be considerably higher, because of the many  terms involved in the cell. On the other hand, the computation of the gradient for each of these terms could be parallelized, making this approach in principle feasible.


\section{Summary and conclusion}
In this paper we have introduced and benchmarked the LC-CTMRG approach to accurately compute the energy variance for iPEPS. It is based on evaluating correlators between Hamiltonian terms in a large cell embedded in the infinite system, where the cell is contracted using  CTMRG. The method thus naturally combines existing iPEPS techniques, making its implementation relatively straightforward. We have demonstrated that the variance converges much more rapidly as a function of the contraction dimension $\chi$ than in a previous approach based on $H$-environments~\cite{corboz16b}, enabling precise estimates of the variance at substantially larger bond dimensions~$D$. 

We have tested the approach for the Heisenberg model and for an exactly solvable fermionic system. The energy extrapolated in the variance was found to be in agreement with the QMC and exact results, respectively, providing evidence for the usefulness of the variance extrapolation for iPEPS. A recent application of the approach to the Shastry-Sutherland model helped to obtain an accurate estimate of the first-order phase transition between the plaquette and quantum spin-liquid phases~\cite{corboz25}.

Finally, we have explained how the LC-CTMRG approach can be used to compute $\langle \langle \cal{L^\dagger L} \rangle \rangle$ for open quantum systems to measure the closeness to the steady state. We have shown how this quantity can be used to detect a first-order phase transition, with the dissipative quantum Ising model as an example. We anticipate that this approach will serve as a practical tool in future studies of phase diagrams of  open quantum systems,  to assess the quality of iPEPS representations of steady states, and to study the evolution toward them.

\acknowledgments
This project has received funding from the European Research Council (ERC) under the European Union's Horizon 2020 research and innovation programme (grant agreement No. 101001604).

\bibliography{refs.bib}

\end{document}